\documentclass[12pt,letterpaper]{article}
\pdfoutput=1
\usepackage[utf8]{inputenc} 
\usepackage{amsmath,amssymb,calc, amsthm,bbm, epsfig,psfrag, mathtools}
\usepackage{graphicx, enumerate}
\usepackage[numbers,sort&compress]{natbib}
\usepackage{graphicx}
\usepackage{dsfont}
\usepackage[font=small,labelfont=bf]{caption}
\usepackage{MnSymbol}
\usepackage[title]{appendix}
\usepackage[english]{babel}
\usepackage{simpler-wick}

\usepackage{tensor}

\numberwithin{equation}{section}
\newcommand{\Comment}[1]{{}}
\newcommand{\efill}{\;\;\;\;\;\;\;\;}

\definecolor{darkblue}{rgb}{0,0,139}

\DeclareRobustCommand{\rchi}{{\mathpalette\irchi\relax}}
\newcommand{\irchi}[2]{\raisebox{\depth}{$#1\chi$}}
\usepackage[linktocpage=true]{hyperref}
\hypersetup{
colorlinks=true,
citecolor=darkblue,
linkcolor=darkblue,
urlcolor=darkblue,
pdfauthor={},
pdftitle={},
pdfsubject={}
}
\makeatletter
\renewcommand\section{\@startsection {section}{1}{\z@}
                                   {-3.5ex \@plus -1ex \@minus -.2ex}
                                   {2.3ex \@plus.2ex}
                                   {\normalfont\large\bfseries}}
\renewcommand\subsection{\@startsection{subsection}{2}{\z@}
                                     {-3.25ex\@plus -1ex \@minus -.2ex}
                                     {1.5ex \@plus .2ex}
                                     {\normalfont\bfseries}}
\makeatother

\newcommand{\be}{\begin{equation}}
\newcommand{\ee}{\end{equation}}
\newcommand{\bea}{\begin{eqnarray}}
\newcommand{\eea}{\end{eqnarray}}

\newcommand{\al}{\alpha}
\renewcommand{\d}{\delta}
\newcommand{\e}{\epsilon}

\newcommand{\g}{\gamma}

\newcommand{\s}{\sigma}
\renewcommand{\t}{\theta}

\newcommand{\hlf}{\frac{1}{2}}

\newcommand{\non}{\nonumber}

\newcommand{\p}{\partial}

\newcommand{\rr}{\rightarrow}

\newcommand{\Z}{\mathbb{Z}}

\newcommand{\lp}{\left(}
\newcommand{\rp}{\right)}
\newcommand{\ls}{\left[}
\newcommand{\rs}{\right]}

\newcommand{\bz}{\bar{z}}

\newcommand{\ov}[1]{{\overline{#1}}}

\begin{document}
\begin{titlepage}

\begin{center}
\hfill         \phantom{xxx}

\vskip 2 cm {\Large \bf Constraining boundary conditions in non-rational CFTs} 
\vskip 1.25 cm {\bf Yucong Cai, Daniel Robbins, and Hassaan Saleem\non\\
\vskip 0.2 cm
 
{\it Department of Physics, University at Albany, Albany, NY, USA}

\vskip 0.2 cm
}
\end{center}
\vskip 1.5 cm
\begin{abstract}
We revisit the question of conformal boundary conditions in the compact free boson CFT in two dimensions.  Besides the well-known Neumann and Dirichlet cases, there is an additional proposed one-parameter family of boundary states when the radius is an irrational multiple of the self-dual radius.  These additional states have a continuous open string spectrum, and we give an explicit formula for the density of states.  We also discuss several pathologies of these states, including the possible violation of the cluster condition, and that they have a divergent g-function. 

\baselineskip=18pt
\end{abstract}
\end{titlepage}
\tableofcontents
\newpage
\section{Introduction}
Quantum field theories on spacetimes with boundaries have long been a subject of interest, both to the high energy and condensed matter communities.  As is often the case, by specializing to the class of 2D conformal field theories one can formulate the issues quite precisely~\cite{Cardy:1984bb}.  Let's take the boundary to be along the real axis.  At the boundary, some boundary conditions for our fields must be chosen.  In the interests of preserving some amount of conformal symmetry, we should choose a conformal boundary condition, meaning that the holomorphic and antiholomorphic stress tensors $T(z)$ and $\ov{T}(\bar{z})$ agree at the boundary $z=\bar{z}$ when we impose our chosen boundary conditions.  

In this conformal context, we can encode the boundary conditions in a so-called boundary state $|B\rrangle$ which must satisfy various consistency conditions~\cite{Cardy:1989ir, Cardy:1991tv, Lewellen:1991}.  There is a reasonably well-formulated procedure to find, in principle, all of the allowed boundary states.  One first constructs the set of Ishibashi states~\cite{Ishibashi:1988kg}, which formally provide a basis for potential conformal boundary states, and then one attempts to impose the additional constraints (the Cardy condition, the cluster condition) to find the specific linear combinations of Ishibashi states which form consistent boundary states.

For rational conformal field theories, this procedure can be implemented very explicitly to find the set of boundary states which preserve (a diagonal copy of) the theory's chiral algebra.  For irrational theories (or for boundary states in rational theories that only preserves a subalgebra), the story can be considerably more complicated.  

In some sense the simplest example of such a theory is the compact free boson.  Two classes of boundary states have long been known, corresponding to either Dirichlet or Neumann boundary conditions for the scalar field.  The corresponding boundary states come in compact one-parameter families (the boundary value of the boson in the Dirichlet case, and the boundary value of a dual field in the Neumann case).  These states are well-behaved and satisfy all of the consistency conditions.  One can ask about the spectrum of states on the interval sandwiched between a pair of these boundaries (which could be both Dirichlet, both Neumann, or one of each), and one finds a nice discrete spectrum of states.  If we have the same boundary condition on each end of the interval, there is a unique vacuum state and then a gap to the first excited state.

However, somewhat surprisingly, it was recognized by Friedan~\cite{Friedan:99}, and then later explored further by Janik~\cite{Janik:01} and Gaberdiel and Recknagel~\cite{GaberdielRecknagel:2001,Gaberdiel(level1):2001}, that there was an additional one-parameter family of boundary states which seemed to satisfy at least those constraints that were feasible to check.  These boundary states, which we call Friedan-Janik states, have various pathologies.  First among them is that these states cannot correspond in any nice reasonable way to boundary conditions on our scalar field, since these are already exhausted by the Dirichlet and Neumann possibilities.  Moreover, these states can be shown to have a continuous spectrum of states on the interval (and hence, by the state-operator correspondence, a continuous spectrum of boundary operators).  One runs into contradictions if one naively tries to check the cluster condition in the presence of one of these boundaries.  And finally, and likely the origin of the above difficulties, the $g$ function of these boundary states diverges, indicating an infinite number of degrees of freedom localized at such a boundary.

The purpose of this article is to explore these states in some more detail and discuss the pathologies mentioned above.  The fact that such states arise already in the compact free boson is likely an indication that they will arise ubiquitously in non-rational conformal field theories in two dimensions, so we view the exercise of exploring these particular examples to be a valuable one.

The organization of the paper is as follows.  In section~\ref{sec:BCFTReview} we review and establish our notation for boundary conformal field theory in general, and review the compact free boson as the particular example we focus on.  Section~\ref{sec:DensityOfStates} contains our computation of the continuous density of states on the interval between Friedan-Janik boundaries.  Section~\ref{sec:Pathologies} then discusses the (naive) violation of the cluster condition, as well as the arguments that the $g$ function diverges.  Finally, section~\ref{sec:Conclusions} gives our conclusions and some future directions.
\section{Review of BCFTs and the compact free boson}
\label{sec:BCFTReview}

\subsection{Setup}

We will be working in the context of two-dimensional boundary conformal field theory (BCFT), i.e.~a unitary 2D CFT in the presence of a (possibly disconnected) boundary. We will mostly work in the upper half plane with coordinates $z=x+iy$. Occasionally we will also consider an annulus, with a pair of boundaries.

Another coordinate system that we can use puts the boundary on the unit disc. To go to this coordinate system, we do the following conformal transformation from the upper half plane (with coordinate $z$) to the exterior of the unit disc (with coordinate $u$)
\be\label{upper half tp disc}
u=\frac{z+i}{i-z}.
\ee
Lastly, we will also use cylinder coordinates i.e. $\tau$ and $\s$ that make the complex coordinate $w=\tau+i\s$. The relation between these coordinates and the coordinate $z$ on the complex plane is $z=e^{w}$.
Along the boundary, the condition that no energy flows in or out imposes a condition on the stress tensor.  In the upper half plane, the condition is that $T_{xy}=0$ along the real axis $y=0$.  In terms of complex coordinates, this condition is
\begin{equation}\label{bd cond on stress}
T(z)=\overline{T}(\bar{z}),\qquad\mathrm{for\ }\bar{z}=z.
\end{equation}
The boundary condition is encoded in a ``boundary state'' $||B\rrangle$. The condition \eqref{bd cond on stress} can now be stated in terms of the Virasoro modes as follows,
\begin{equation}\label{bd cond on virasoro}
\lp L_n-\overline{L}_{-n}\rp||B\rrangle=0,\qquad\forall n\in\Z.
\end{equation}
These boundary states are built of Ishibashi states $||i\rrangle$ for different primary states $|i\rangle$ \cite{Ishibashi:1988kg}. The Ishibashi state $||i\rrangle$ is the unique state that is built by acting on $|i\rangle$ with oscillators and it satisfies \eqref{bd cond on virasoro} (with $||B\rrangle$ relaced by $||i\rrangle$).

A general boundary state $||B_{\al}\rrangle$ in terms of Ishibashi states is
\be\label{general boundary in ishibashi}
\left|B_\al\right\rrangle=\sum_iA_{\al i}\left|i\right\rrangle,
\ee

We will use the radial quantization Hamiltonian, which is
\be\label{Hamiltonian}
H=\hlf\lp L_{0}+\ov{L}_{0}-\frac{c}{12}\rp,
\ee
\subsection{Consistency conditions}
There are a number of consistency conditions that a CFT needs to satisfy \cite{Cardy:1989ir, Sonoda:1988, Sonoda2:1988, Lewellen:1991} but in this paper, we will mostly be concerned with Cardy condition \cite{Cardy:1989ir} and Cluster condition \cite{Lewellen:1991}. We will describe these conditions briefly.

We can calculate the overlap of two boundary states as follows;
\be\label{boundary overlap closed}
\llangle B_{\al}|q^{H}|B_{\g}\rrangle=\sum_{i,j}\ov{A}_{\al i}A_{\g j}\llangle i|q^{H}|j\rrangle=\sum_i\ov{A}_{\al i}A_{\g i}\chi_{h_{i}}(q).
\ee
where $\chi_{h_{i}}(q)$ is the character of the highest weight representation built on $|i\rangle$, and \mbox{$q^{H}=e^{2\pi i\tau H}$} ($\tau$ is the modular parameter of a torus). We can transform this expression to the open string description by doing the modular $S$ transformation
\be\label{s transformation beta q}
q\rr \tilde{q}=e^{-\frac{2\pi i}{\tau}}.
\ee
In the open string spectrum, the amplitude in \eqref{boundary overlap closed} becomes a partition function and thus, we expect to get an equation of the following form;
\be\label{cardy condition essence}
\llangle B_{\al}|q^{H}|B_{\g}\rrangle=\sum_{j}n_{\al\g}^{j}\chi_{h_{j}}(\tilde{q}).
\ee
Since the amplitude in \eqref{boundary overlap closed} becomes a partition function under the S transformation, the coefficient $n_{\al\g}^{j}$ is counting the number of times the character $\chi_{h_{j}}(\tilde{q})$ is contributing to the partition function. Therefore, all $n_{\al\g}^{j}$ should be non-negative integers i.e. $n_{\al\g}^{j}\in\Z_{\ge 0}$ .
Moreover, when $\al=\g$, we expect that the vacuum state should appear once, $n_{\al\al}^0=1$.  These requirements are the essence of a nontrivial condition on boundary states \eqref{general boundary in ishibashi} called the \textit{Cardy condition}.

The cluster condition is a special case of a sewing constraint in \cite{Lewellen:1991}. Consider two operators $\phi_{i}$ and $\phi_{j}$ at $z=iy$ and $z=x+iy$ respectively. The cluster condition arises in the special limit $x\rr\infty$ holding $y$ fixed. In terms of the boundary states, the cluster condition can be written as follows,
\be\label{cluster condition}
\lim_{x\rr \infty}\langle 0| \phi_{i}(0,y)\phi_{j}(x,y)|B\rrangle=\lim_{x\rr \infty}\langle 0| \phi_{i}(0,y)|B\rrangle \langle0|\phi_{j}(x,y)|B\rrangle.
\ee 
This condition can be written in the $u$ coordinate as well (see \eqref{Dirichlet cluster u}). For rational CFTs (RCFTs) we can write the cluster condition in a more convenient form. To write this form, we define relative coefficients from \eqref{general boundary in ishibashi},
$$
B_{\al i}=\frac{A_{\al i}}{A_{\al 0}}.
$$
The cluster condition then constrains these coefficients~\cite{Fuchs:1997kt},
\be\label{cluster for RCFT}
B_{\al i}B_{\al j}=\sum_{k}M_{ij}^{\;\;k}B_{\al k},
\ee
where the $M_{ij}^{\;\;k}$ coefficients are defined in terms of OPE coefficients $C_{ij}^{\;\;k}$ and fusion matrices $F_{k0}$,
\be\label{m coefficients def}
M_{ij}^{\;\;k}=C_{ij}^{\;\;k}F_{k0}\ls\begin{matrix}
j&j\\
i&i
\end{matrix}
\rs,
\ee
with $0$ denoting the identity operator.
\subsection{The compact free boson CFT}
In this subsection, we will give a brief review of the facts about the compact free boson CFT which are relevant for this paper. We set the self dual radius equal to $1$ throughout this paper.  The compact free boson $X$ with radius $R$ has the following identification,
\be
X\sim X+2\pi R,
\ee
and the following mode expansion (with $X=X_{L}(z)+X_{R}(\bz)$)
\begin{align}
\label{eq:XLModes}
    X_L(z)=\ & \widehat{x}_{L0}-\frac{i}{2}\lp\frac{\widehat{N}}{R}+\widehat{M}R\rp\ln z+\frac{i}{\sqrt{2}}\sum_{n\ne 0}\frac{1}{n}\al_nz^{-n},\\
    \label{eq:XRModes}
    X_R(\bar{z})=\ & \widehat{x}_{R0}-\frac{i}{2}\lp\frac{\widehat{N}}{R}-\widehat{M}R\rp\ln\bar{z}+\frac{i}{\sqrt{2}}\sum_{n\ne 0}\frac{1}{n}\widetilde{\al}_n\bar{z}^{\,-n}.
\end{align}
The operators $\widehat{N}$ and $\widehat{M}$ are momentum and winding operators that have been normalized to have integer eigenvalues. Going forward, for $n\neq 0$, we will use the following oscillators instead of the $\al_{n}$ and $\tilde{\al}_{n}$ oscillators (for $n>0$),
$$
a_{n}=\frac{1}{\sqrt{n}}\al_{n},\efill a^{\dagger}_{n}=\frac{1}{\sqrt{n}}\al_{-n}
$$
and similarly for $\tilde{a}_{n}$ and $\tilde{a}^{\dagger}_{n}$. The nonvanishing commutation relations are
\begin{equation}
\ls a_n,a_m^\dagger\rs=\ls\widetilde{a}_n,\widetilde{a}_m^\dagger\rs=\d_{n+m,0},\qquad\ls\widehat{x}_0,\widehat{N}\rs=iR,\qquad\ls\widehat{\tilde{x}}_0,\widehat{M}\rs=\frac{i}{R}.
\end{equation}
where 
$$
\widehat{x}_{0}=\widehat{x}_{L0}+\widehat{x}_{R0}\efill\efill \widehat{\tilde{x}}_{0}=\widehat{x}_{L0}-\widehat{x}_{R0}
$$
To calculate the Hamiltonian, we will need the expression for $L_{0}$ which is given as follows for this theory
\begin{align}
L_0=& \frac{1}{4}\lp\frac{\widehat{N}}{R}+\widehat{M}R\rp^2+\sum_{n=1}^\infty na^{\dagger}_na_n
\end{align}
Using the OPE $$
\p X(z)\p X(0)\sim -\hlf z^{-2},
$$
we can check that this theory has central charge $c=\bar{c}=1$. We can think of the compact free boson as a $U(1)$ current algebra with currents
$$
J(z)=\sqrt{2}i\p X_L(z)\efill \bar{J}(z)=\sqrt{2}i\p X_R(\bz)
$$
In this language, the states $|(N,M)\rangle$ are the primary states of the current algebra. Under the state operator correspondence, the $U(1)$ primary states $|(N,M)\rangle$ map to (normal ordered) exponential operators
\be
\label{eq:NMExpOps}
\mathcal{V}_{(N,M)}(z,\bar{z})=:\exp\ls i\lp\frac{N}{R}+MR\rp X_L(z)+i\lp\frac{N}{R}-MR\rp X_R(\bar{z})\rs:.
\ee
with the following OPE,
\be
\mathcal{V}_{(N,M)}(z,\bar{z})\mathcal{V}_{(N',M')}(0,0)\sim z^{\hlf\lp\frac{N}{R}+MR\rp\lp\frac{N'}{R}+M'R\rp}\bz^{\hlf\lp\frac{N}{R}-MR\rp\lp\frac{N'}{R}-M'R\rp}\mathcal{V}_{(N+N',M+M')}(0,0)
\ee
Moreover, the $U(1)$ characters $\chi_{h}(q)$ break further into Virasoro characters when $h$ is the square of an integer or a half integer i.e. $\chi_h^{U(1)}(q)=\chi_h(q)$ if $2\sqrt{h}\nin\Z$, and 
\be
\chi_{J^2}^{U(1)}(q)=\sum_{k=0}^\infty\chi_{(J+k)^2}(q),
\ee
where $J$ is a non-negative integer or half-integer.

This means that if $\frac{N}{R}\pm MR\in\Z$ for some integers $N$ and $M$, then the Fock space built on the corresponding state $|(N,M)\rangle$ will split into degenerate representations.  For a given choice of $R$, the full set of Virasoro primaries will be the states $|(N,M)\rangle$ along with the series of degenerate representations that come along with every solution to $\frac{N}{R}\pm MR\in\Z$.

In this section and the next we will mostly restrict to the case that $R$ is a sufficiently generic irrational multiple of the self-dual radius so that the only choice of $N$ and $M$ which leads to degeneracies is $N=M=0$.  We will label the primaries that are current algebra descendants of the vacuum by $|[J,J']\rangle$, where $J,J'=0,1,2,\cdots$.  The full set of Virasoro primaries in this theory then consists of the states given in Table 1.
\begin{table}[t]
    \centering
    \begin{tabular}{c|c|c|c}
         \textbf{State}& \textbf{Conditions}& $\mathbf{h}$ & $\mathbf{\overline{h}}$\\
         \hline
         $|(N,M)\rangle$&$N,M\in\Z/\{0\}$&$h=\frac{1}{4}\lp\frac{N}{R}+MR\rp^2$&$\bar{h}=\frac{1}{4}\lp\frac{N}{R}-MR\rp^2$\\
$|[J,J']\rangle$&$J,J'\in\Z_{\geq 0}$&$h=J^2$&$\bar{h}=J'^2$
    \end{tabular}
    \caption{Virasoro primaries for the free compact boson with irrational $R^{2}$}
    \label{tab:my_label}
\end{table}
Under the state operator correspondence, the $|[J,J']\rangle$ states correspond to the operators $V_{[J,J']}(z,\bz)$ which can be written as 
$$
V_{[J,J']}(z,\bz)=\mathcal{N}_{JJ'}U_J(z)\ov{U}_{J'}(\bar{z})
$$
where $\mathcal{N}_{JJ'}$ is a normalization constant whose details won't be important, and the operators $U_J(z)$ can be formally defined as
\be
\label{eq:UJOps}
U_J(z)=\lp\oint\frac{du}{2\pi}:e^{-2iX_L(u+z)}:\rp^J:e^{2iJX_L(z)}:,
\ee
with a similar expression for $\ov{U}_{J'}(\bar{z})$.  Note that the individual exponentials in this expression are not well-quantized operators at generic $R$ values, but they do make sense at the self-dual radius, $R=1$. Expressions for the operators $U_{J}(z), \ov{U}_{J'}(\bar{z})$ and the states $||[J,J]\rrangle$ can be written in terms of Schur polynomials \cite{kac2013bombay, Kac:1978ge, segal1981unitary, wakimoto1986fock}. After taking the OPEs in \eqref{eq:UJOps}, the resulting normal-ordered operator is built only from derivatives of $X_L$, and can then be interpreted at any value of the radius $R$.  We are essentially using the $SU(2)$ current symmetry that is present at the self-dual radius to construct the operators there, by acting on a highest weight state with lowering operators and appealing to the fact that, while the intermediate states are not well-defined at generic $R$, the $m_J=0$ state is well-defined and is $R$-independent. 

There are two well-known boundary states in the free boson theory i.e. the Dirichlet and Neumann states. These states are written as follows
\be\label{DirichletinIshibashi}
||D(x_{0})\rrangle=\frac{1}{\sqrt{\sqrt{2}R}}\sum_{N\in\Z}e^{-\frac{iN}{R}x_{0}}||(N,0)\rrangle,
\ee
\be\label{NeumanninIshibashi}
||N(\tilde{x}_{0})\rrangle=\sqrt{\frac{R}{\sqrt{2}}}\sum_{M\in\Z}e^{iMR\tilde{x}_{0}}||(0,M)\rrangle.
\ee
These states can easily be shown to satisfy the Cardy condition (for example, see \cite{Becker(orbifold):2017}) and they also satisfy the cluster condition. For instance, to show that the Dirichlet states satisfy the cluster condition, one can write down the cluster condition in the $u$ coordinates (defined in \eqref{upper half tp disc}) as follows
\be\label{Dirichlet cluster u}
\lim_{u_{2}\rightarrow -1}\langle 0|\phi_{n}(u_{1})\phi_{n'}(u_{2})|D(x_{0})\rrangle=\lim_{u_{2}\rightarrow -1}\langle 0|\phi_{n}(u_{1})|D(x_{0})\rrangle\langle 0|\phi_{n'}(u_{2})|D(x_{0})\rrangle.
\ee
where $\phi_{n}(u_{j})=e^{inX(u_{j})/R}$, and then calculate the following quantities
\be\label{one point corr}
\langle0|:e^{inX_{1}/R}:|D(x_{0})\rrangle\langle0|:e^{in'X_{2}/R}:|D(x_{0})\rrangle=(|u_{1}|^{2}-1)^{-\lp \frac{n}{R}\rp^{2}}(|u_{2}|^{2}-1)^{-\lp \frac{n'}{R}\rp^{2}},
\ee
\be\label{two point corr}
\langle0|:e^{inX_{1}/R}::e^{in'X_{2}/R}:|D(x_{0})\rrangle=(|u_{1}|^{2}-1)^{-\lp \frac{n}{R}\rp^{2}}(|u_{2}|^{2}-1)^{-\lp \frac{n'}{R}\rp^{2}}|u_{1}-u_{2}|^{ \frac{2nn'}{R^{2}}}|u_{1}\bar{u}_{2}-1|^{- \frac{2nn'}{R^{2}}}.
\ee
where $X_{i}$ is a shorthand for $X(u_{i},\bar{u}_{i})$. We see that in the limit $u_{2}\rightarrow -1$, then the right-hand sides of both \eqref{one point corr} and \eqref{two point corr} have identical leading behavior, and thus \eqref{Dirichlet cluster u} is satisfied. One can do a similar calculation with other bulk operators or with Neumann states.
\subsection{Friedan-Janik states}
It was pointed out by Friedan \cite{Friedan:99} that at irrational values of $R^{2}/R^{2}_{\text{self-dual}}$, there is a continuous spectrum of boundary states, labeled by $\theta$ (with $0<\theta<\pi$). These boundary states were worked out in \cite{Janik:01} by imposing some of the cluster conditions \eqref{cluster for RCFT}. What was shown in \cite{Janik:01} is that if we consider the following boundary state;
$$
||B\rrangle=\sum_{J=0}^{\infty}A_{J}||[J,J]\rrangle,
$$
then \eqref{cluster for RCFT} can be satisfied if $i,j$ and $k$ run over the $|[J,J]\rangle$ primaries. Moreover, it was shown that \eqref{cluster for RCFT} is satisfied only if the following recurrence relation is satisfied
$$
B_{J}\cos{\theta}=\frac{J}{2J+1}B_{J-1}+\frac{J+1}{2J+1}B_{J+1}\;\;\;\text{ where }B_{J}=\frac{A_{J}}{A_{0}},\;\;B_{1}=\cos{\theta}
$$
This recurrence relation sets $B_{J}=P_{J}(\cos{\theta})$ where $P_{J}(\cos{\theta})$ are Legendre functions (the $\cos{\theta}$ dependence just shows that all $B_{J}$'s - except $B_{0}$, which is 1 - are dependent on $B_{1}$). The form of these boundary states (which we will call the Friedan-Janik boundary states) is
\be\label{Friedanstate}
||F(\cos{\theta})\rrangle=\mathcal{C}(\theta)\sum_{J=0}^{\infty}P_{J}(\cos{\theta})||[J,J]\rrangle
\ee
where $\theta\in(0,\pi)$, and $\mathcal{C}(\theta)$ is a possible normalization constant.
\section{Density of states for Friedan-Janik boundaries}
\label{sec:DensityOfStates}
The overlap of two Friedan-Janik states and its form in the open-string sector was calculated in \cite{Janik:01} and the result was given in a double integral form, which we reproduce here
\be\label{Friedanopensectormain}
\llangle F(\cos{\theta_{1}})|q^{H}|F(\cos{\theta_{2}})\rrangle=\frac{\overline{\mathcal{C}({\theta}_{1})}\mathcal{C}({\theta}_{2})}{\sqrt{2}\pi^{2}}\int_{0}^{\pi}d\psi\int_{0}^{\pi}d\phi\sum_{n\in\Z}\rchi_{\frac{1}{4}\left(n-\frac{t}{2\pi}\right)^{2}}(\tilde{q})
\ee
where the relation of $t$ in terms of $\theta_{1}$, $\theta_{2}$, $\phi$ and $\psi$ is given by the following set of equations;
$$
\cos{\frac{t}{2}}=\cos{\frac{\theta}{2}}\cos{\frac{\phi}{2}}
$$
\be\label{t relations}
\cos{\theta}=\cos{\theta_{1}}\cos{\theta_{2}}-\sin{\theta_{1}}\sin{\theta_{2}}\cos{\psi}
\ee
This overlap should represent the partition function for open strings stretched between boundaries $|F(\cos\theta_1)\rrangle$ and $|F(\cos\theta_2)\rrangle$.  The fact that it is written as an integral rather than a sum likely indicates a continuous, rather than discrete, spectrum.  This would seem to be in tension with the more usual form of the Cardy condition, in which the coefficients of characters in this partition function should be non-negative integers.  Note that we do at least have a positive integrand, so there should be a well-defined density of states $\rho(h)$.  To explore the properties of this result in more detail, we will try to write \eqref{Friedanopensectormain} as
\be\label{Friedan desired form}
\llangle F(\cos\t_{1})|q^{H}|F(\cos\t_{2})\rrangle=\int_{0}^{\infty}dh\;\rho(h)\chi_{h}(\tilde{q})
\ee
for some $\rho(h)$.

First we note that as $\psi$ runs from $0$ to $\pi$, $\cos\theta$ runs between $\cos(\theta_1+\theta_2)$ and $\cos(\theta_1-\theta_2)$.  If $\theta_1+\theta_2\le\pi$, then this corresponds to the $\theta$ coordinate running from $|\theta_1-\theta_2|$ to $\theta_1+\theta_2$.  On the other hand, if $\theta_1+\theta_2\ge\pi$, then (under the assumption that $\theta$ remains in the range $0\le\theta\le\pi$) $\theta$ runs from $|\theta_1-\theta_2|$ to $2\pi-\theta_1-\theta_2$.  In fact, the cases don't have to be treated separately since $\theta$ is symmetric under swapping $\theta_1$ and $\theta_2$, and is also unchanged under the simultaneous replacement of $\theta_1$ by $\pi-\theta_1$ and $\theta_2$ by $\pi-\theta_2$, so without loss of generality we will assume that $\theta_1\le\theta_2$ and that $\theta_1+\theta_2\le\pi$.  For fixed $\theta$, $t$ ranges from $\theta$ up to $\pi$ as $\phi$ varies.
One immediate observation is that if $\theta_1<\theta_2$ we have $t\ge\theta\ge\theta_2-\theta_1>0$.  Then since the conformal weight $h$ is related to $t$ (and some integer $n$) via
\be
h=\frac{1}{4}\lp n-\frac{t}{2\pi}\rp^2,\qquad n\in\Z,
\ee
we see that in such a sector the conformal weights are bounded away from the squares of half-integers.  The spectrum would seem to have a banded structure, with no states in a neighborhood of each square of a half-integer, meaning in particular that there are no degenerate representations appearing.

If we change variables from $\psi$ and $\phi$ to $\theta$ and $t$, we use 
\begin{align}
\left|d\phi\,d\psi\right|=\ & \frac{\sin\frac{t}{2}}{\cos\frac{\theta}{2}}\lp 1-\frac{\cos^2\frac{t}{2}}{\cos^2\frac{\theta}{2}}\rp^{-\hlf}\left|dt\,d\psi\right|\\
=\ & \sin\theta\sin\frac{t}{2}\lp\cos^2\frac{\theta}{2}-\cos^2\frac{t}{2}\rp^{-\hlf}\lp\sin^2\theta_1\sin^2\theta_2-\lp\cos\theta_1\cos\theta_2-\cos\theta\rp^2\rp^{-\hlf}\left|dt\,d\theta\right|\non\\
=\ & \frac{\sqrt{2}\sin\theta\sin\frac{t}{2}\ \left|dt\,d\theta\right|}{\sqrt{\lp\cos\theta-\cos t\rp\lp\cos\theta-\cos(\theta_1+\theta_2)\rp\lp\cos(\theta_2-\theta_1)-\cos\theta\rp}}.\non
\end{align}

Let's define
\be
f(\theta,t)=\frac{\sin\theta}{\sqrt{\lp\cos\theta-\cos t\rp\lp\cos\theta-\cos(\theta_1+\theta_2)\rp\lp\cos(\theta_2-\theta_1)-\cos\theta\rp}}.
\ee
Then taking the integration region into account, we can write
\begin{align}
\llangle F(\cos{\theta_{1}})|q^{H}|F(\cos{\theta_{2}})\rrangle=\ & \frac{\ov{\mathcal{C}(\t_{1})}\mathcal{C}(\t_{2})}{\pi^2\eta(\tilde{q})}\sum_{n\in\Z}\lp
\int_{\theta_2-\theta_1}^{\theta_1+\theta_2} dt\,\sin\frac{t}{2}\,\tilde{q}^{\,\frac{1}{4}\lp n-\frac{t}{2\pi}\rp^2}
\int_{\theta_2-\theta_1}^t
f(\theta,t)\,d\theta\right.\non\\
& \qquad\left. 
+\int_{\theta_1+\theta_2}^\pi dt\,\sin\frac{t}{2}\,\tilde{q}^{\,\frac{1}{4}\lp n-\frac{t}{2\pi}\rp^2}\int_{\theta_2-\theta_1}^{\theta_1+\theta_2}
f(\theta,t)\,d\theta\rp.
\end{align}
Using
\be
\left|dt\right|=2\pi\frac{\left|dh\right|}{\sqrt{h}},
\ee
we can extract the density of states.  For $h<0$ or for
\be
\frac{1}{4}\lp n-\frac{\theta_2-\theta_1}{2\pi}\rp^2<h<\frac{1}{4}\lp n+\frac{\theta_2-\theta_1}{2\pi}\rp^2,\qquad n\ge 0,
\ee
we have $\rho(h)=0$, and otherwise
\be
\rho(h)=2\ov{\mathcal{C}(\t_{1})}\mathcal{C}(\t_{2})\frac{\left|\sin(2\pi\sqrt{h})\right|}{\pi\sqrt{h}}\int_{\theta_2-\theta_1}^{\Theta(h)}f(\theta,4\pi\sqrt{h})\,d\theta,
\ee
where the upper bound on the integration is given by
\be
\Theta(h)=\left\{\begin{matrix} 2\pi\lp 2\sqrt{h}-n\rp, & \frac{1}{4}\lp n+\frac{\theta_2-\theta_1}{2\pi}\rp^2\le h<\frac{1}{4}\lp n+\frac{\theta_1+\theta_2}{2\pi}\rp^2, \\ \theta_1+\theta_2, & \frac{1}{4}\lp n+\frac{\theta_1+\theta_2}{2\pi}\rp^2\le h\le\frac{1}{4}\lp n+1-\frac{\theta_1+\theta_2}{2\pi}\rp^2, \\ 2\pi\lp n+1-2\sqrt{h}\rp, & \frac{1}{4}\lp n+1-\frac{\theta_1+\theta_2}{2\pi}\rp^2<h\le\frac{1}{4}\lp n+1-\frac{\theta_2-\theta_1}{2\pi}\rp^2,\end{matrix}\right.
\ee
where $n\ge 0$ is an integer.

Actually, we can push even further than this.  In each of the intervals with nonzero $\rho(h)$, the $\theta$ integral has the form
\be
\int_a^b\frac{\sin\theta\,d\theta}{\sqrt{\lp\cos\theta-\cos b\rp\lp\cos\theta-\cos c\rp\lp\cos a-\cos\theta\rp}},
\ee
where $a\le b\le c$.  By making a change of variables
\be
u^2=\frac{\cos\theta-\cos b}{\cos a-\cos b},
\ee
the integral becomes
\be
\frac{2}{\sqrt{\cos b-\cos c}}\int_0^1\frac{du}{\sqrt{\lp 1-u^2\rp\lp 1+\g u^2\rp}}=\frac{2}{\sqrt{\cos b-\cos c}}K(-\g),
\ee
where
\be
\g=\frac{\cos a-\cos b}{\cos b-\cos c},
\ee
and where $K(m)=K(k^2)$ is the complete elliptic integral of the first kind, defined by
\be
K(m)=\int_0^1\frac{du}{\sqrt{\lp 1-u^2\rp\lp 1-mu^2\rp}}.
\ee
$K(m)$ is most commonly defined for real values of $m$ between $0$ and $1$, but the integral is also well-defined and convergent for negative values of $m$ (and indeed everywhere on the complex plane except for a branch cut running along the real axis from $m=1$ to $m=+\infty$).  On the negative real axis, $K(m)$ is real and positive.  At zero we have $K(0)=\pi/2$, and for large positive $\g$ the leading behavior is $K(-\g)\cong\ln(\g)/2\sqrt{\g}$.

\begin{figure}
\centering
\includegraphics[scale=0.42]{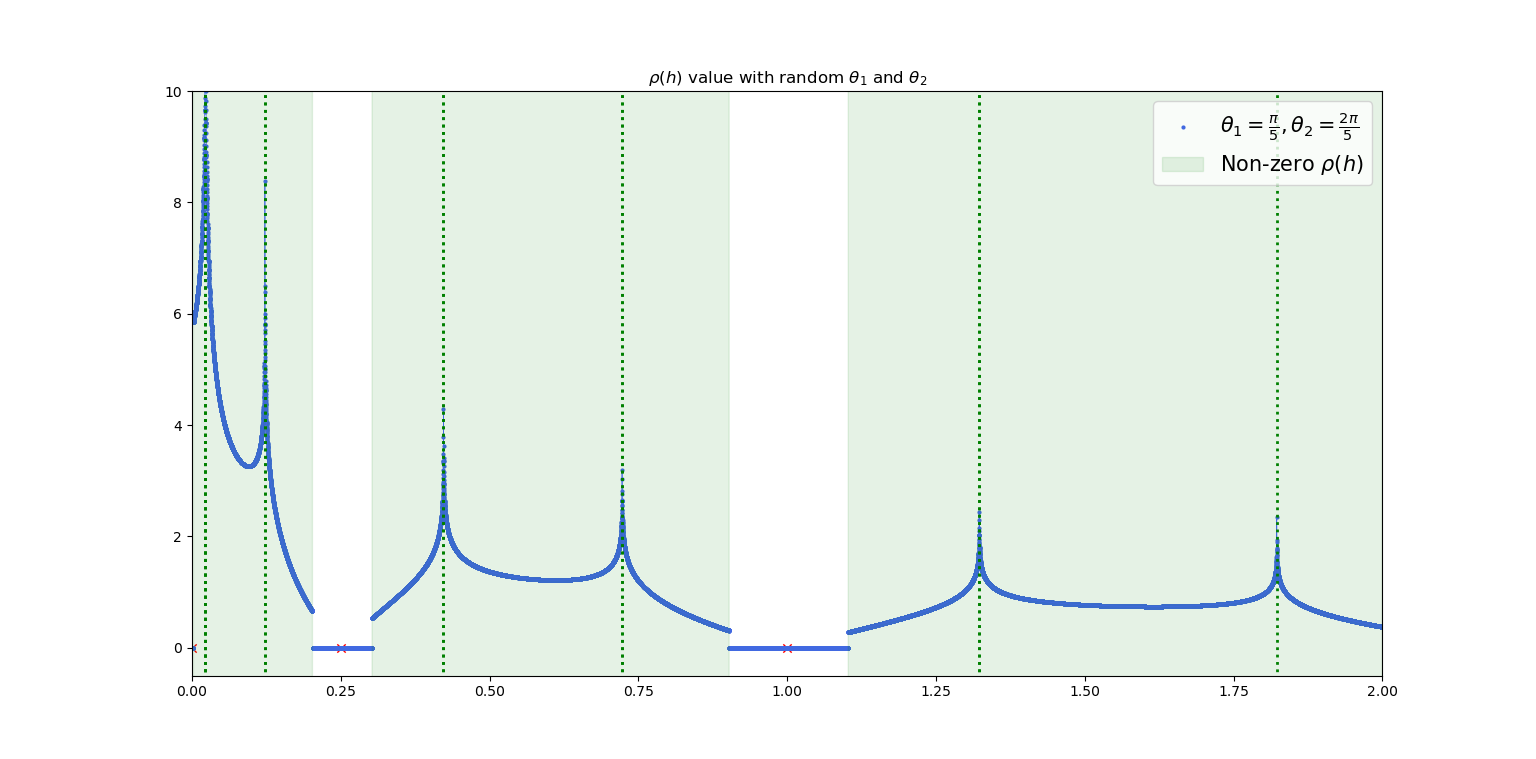}
\caption{A typical example of the density of states with a clear band structure.  The white regions indicate gaps in the spectrum, while in the shaded regions, $\rho(h)$ indicates a continuous spectrum.  Though it's difficult to see, there is an initial gap from $h=0$ to $h=0.0025$.  The dashed lines indicate divergences at $\frac{1}{4}(n\pm 0.3)^2$.}
\label{fig:DoST}
\end{figure}

Thus our final result for the density of states is, for each $n\ge 0$,
\be
\rho(h)=\left\{\begin{matrix} 0, & \mathrm{region\ }\text{I}_n, \\ \frac{2\sqrt{2}\,\ov{\mathcal{C}(\t_{1})}\mathcal{C}(\t_{2})}{\pi\sqrt{h}}\sqrt{\frac{1-\cos(4\pi\sqrt{h})}{\cos(4\pi\sqrt{h})-\cos(\theta_1+\theta_2)}}K(-\frac{\cos(\theta_2-\theta_1)-\cos(4\pi\sqrt{h})}{\cos(4\pi\sqrt{h})-\cos(\theta_1+\theta_2)}), & \mathrm{region\ }\text{II}_n, \\ \frac{2\sqrt{2}\,\ov{\mathcal{C}(\t_{1})}\mathcal{C}(\t_{2})}{\pi\sqrt{h}}\sqrt{\frac{1-\cos(4\pi\sqrt{h})}{\cos(\theta_1+\theta_2)-\cos(4\pi\sqrt{h})}}K(-\frac{\cos(\theta_2-\theta_1)-\cos(\theta_1+\theta_2)}{\cos(\theta_1+\theta_2)-\cos(4\pi\sqrt{h})}), & \mathrm{region\ }\text{III}_n, \end{matrix}\right.
\ee
where the regions are defined by (here $n=\lfloor 2\sqrt{h}\rfloor$ is a non-negative integer)
\begin{align}
    \mathrm{region\ }\text{I}_n: & \ls\left.\frac{1}{4} n^2,\frac{1}{4}\lp n+\frac{\theta_2-\theta_1}{2\pi}\rp^2\rp\right.\cup \ls\left.\frac{1}{4} \lp n+1-\frac{\theta_2-\theta_1}{2\pi}\rp^2,\frac{1}{4}\lp n+1\rp^2\rp\right., \non\\
    \mathrm{region\ }\text{II}_n: & \ls\left.\frac{1}{4}\lp n+\frac{\theta_2-\theta_1}{2\pi}\rp^2,\frac{1}{4}\lp n+\frac{\theta_1+\theta_2}{2\pi}\rp^2\rp\right.\non\\
    & \qquad\cup \ls\left.\frac{1}{4}\lp n+1-\frac{\theta_1+\theta_2}{2\pi}\rp^2,\frac{1}{4}\lp n+1-\frac{\theta_2-\theta_1}{2\pi}\rp^2\rp\right., \non\\
    \mathrm{region\ }\text{III}_n: & \ls\left.\frac{1}{4}\lp n+\frac{\theta_1+\theta_2}{2\pi}\rp^2,\frac{1}{4}\lp n+1-\frac{\theta_1+\theta_2}{2\pi}\rp^2\rp\right.
\end{align}
For generic values with $0<\theta_1<\theta_2<\pi$ and $\theta_1+\theta_2<\pi$, we start with a gap in the spectrum from $h=0$ to $h=(\theta_2-\theta_1)^2/16\pi^2$, then $\rho(h)$ jumps to a finite value of
\be
\label{eq:ValueAtGap}
\rho\lp\frac{\lp\theta_2-\theta_1\rp^2}{16\pi^2}\rp=\frac{4\sqrt{2}\pi\ov{\mathcal{C}(\t_{1})}\mathcal{C}(\t_{2})}{\theta_2-\theta_1}\sqrt{\frac{1-\cos(\theta_2-\theta_1)}{\cos(\theta_2-\theta_1)-\cos(\theta_1+\theta_2)}}.
\ee
After that, $\rho(h)$ increases until $h=h_0=(\theta_1+\theta_2)^2/16\pi^2$ where $\rho(h)$ diverges logarithmically (from both sides),
\be
\rho(h)\cong\frac{4\sqrt{2}\,\ov{\mathcal{C}(\t_{1})}\mathcal{C}(\t_{2})}{\theta_1+\theta_2}\sqrt{\frac{1-\cos(\theta_1+\theta_2)}{\cos(\theta_2-\theta_1)-\cos(\theta_1+\theta_2)}}\ln\left|h-h_0\right|.
\ee
In particular, although $\rho(h)$ diverges at this point, the divergence is integrable, as one would want for a density of states.  Then there is a central band where $\rho(h)$ comes back down from its divergence, reaches a minimum value, and then rises up to diverge logarithmically again at the point $h_0=(2\pi-\theta_1-\theta_2)^2/16\pi^2$.  From there it decreases again to a finite value at the point $h=(2\pi-\theta_2+\theta_1)^2/16\pi^2$ where it discontinuously drops to zero and we enter another gap region.  The gap has a finite width containing the point $h=1/4$, and then the band structure repeats.  Each $h=n^2/4$ lies inside one of the gaps, while each $h=(n+\hlf)^2/4$ lies between a pair of divergences.  A representative example is sketched in Figure~\ref{fig:DoST}.

\begin{figure}
    \centering
    \includegraphics[scale=0.42]{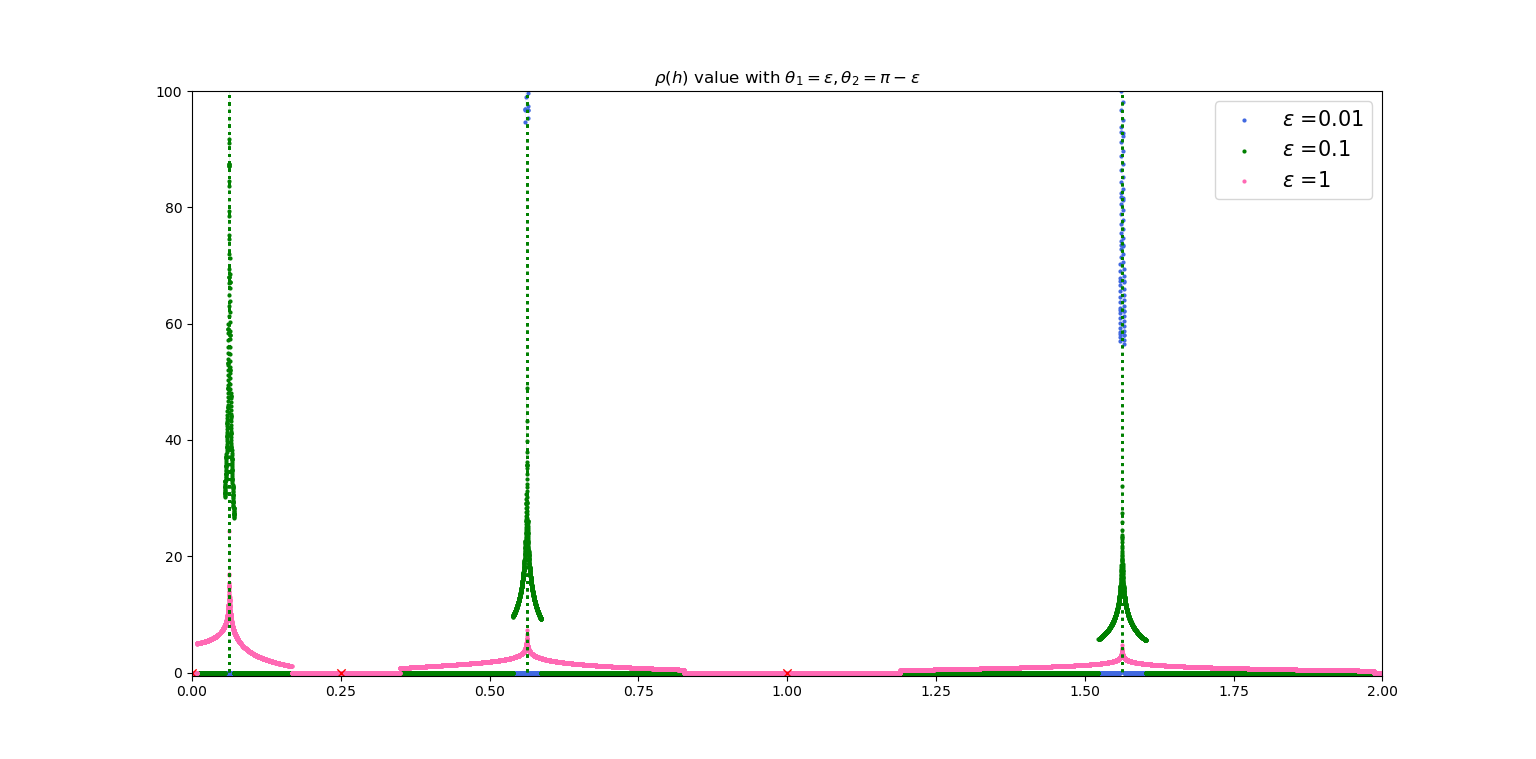}
    \caption{Three examples of the density of states when $\theta_2=\pi-\theta_1$.  As $\theta_1$ gets smaller, the gaps get larger and the bands narrower, and the density of states approaches a sum of delta functions.}
    \label{fig:DoSC}
\end{figure}

Two special cases deserve closer examination.  The first is when $\theta_2=\pi-\theta_1$.  In this case, the two divergent points in each band coalesce into a single divergence at $h=(n+\hlf)^2/4$, and the density of states is
\be
\rho(h)=\left\{\begin{matrix}0 & \frac{1}{4}n^2\le h<\frac{1}{4}\lp n+\hlf-\frac{\theta_1}{\pi}\rp^2 \\ \frac{2\sqrt{2}\,\ov{\mathcal{C}(\t_{1})}\mathcal{C}(\pi-\t_{1})}{\pi\sqrt{h}}\left|\tan(2\pi\sqrt{h})\right|K\lp\frac{\cos(2\theta_1)+\cos(4\pi\sqrt{h})}{1+\cos(4\pi\sqrt{h})}\rp & \frac{1}{4}\lp n+\hlf-\frac{\theta_1}{\pi}\rp^2\le h<\frac{1}{4}\lp n+\hlf+\frac{\theta_1}{\pi}\rp^2 \\ 0 & \frac{1}{4}\lp n+\hlf+\frac{\theta_1}{\pi}\rp^2\le h<\frac{1}{4}\lp n+1\rp^2. \end{matrix}\right.
\ee
Three representative examples are plotted in Figure~\ref{fig:DoSC}.  Of particular interest is the case when $\theta_1=\e\rr 0$ and $\theta_2=\pi-\e\rr\pi$.  In this case the gaps expand to fill almost everywhere, and the bands shrink down to be localized at the points $\frac{1}{4}(n+\hlf)^2$.  Since the density of states is integrable, this means that in the limit we must have a sum of delta functions,
\be
\label{eq:SumOfDeltas}
\lim_{\e\rr 0}\rho(h)=\ov{\mathcal{C}(0)}\mathcal{C}(0)\sum_{n=0}^\infty c_n\d\lp h-\frac{1}{4}\lp n+\hlf\rp^2\rp,
\ee
where the $c_n$ are some constants.  To compute $c_n$, we can compute the integral of $\rho(h)$ from the gap to the divergence,
\be
    \frac{2\sqrt{2}}{\pi}\lim_{\e\rr 0}\int_{\frac{1}{4}\lp n+\hlf-\frac{\e}{\pi}\rp^2}^{\frac{1}{4}\lp n+\hlf\rp^2}\frac{dh}{\sqrt{h}}\tan(2\pi\sqrt{h})K\lp\frac{\cos(2\e)+\cos(4\pi\sqrt{h})}{1+\cos(4\pi\sqrt{h})}\rp.
\ee
Changing variables using $h=\frac{1}{4}\lp n+\hlf-\frac{\e u}{\pi}\rp^2$, $dh/\sqrt{h}=-\e\,du/\pi$, we have
\be
\frac{2\sqrt{2}\e}{\pi^2}\lim_{\e\rr 0}\int_0^1du\,\cot(\e u)K\lp\frac{\cos(2\e)-\cos(2\e u)}{1-\cos(2\e u)}\rp=\frac{2\sqrt{2}}{\pi^2}\int_0^1\frac{du}{u}K\lp-\frac{1-u^2}{u^2}\rp
\ee
Since $K(m)$ enjoys an identity
\be
\frac{1}{u}K\lp-\frac{1-u^2}{u^2}\rp=K\lp1-u^2\rp
\ee
the definite integral is one that appears as in \cite{GaberdielRecknagel:2001}, evaluating to $\pi^2/4$, so the integral above becomes simply $1/\sqrt{2}$.  Similar manipulations give us the integral from the divergence to the next gap is identical, so we conclude that $c_n=\sqrt{2}$, independent of $n$.  This result should be compared to the annulus amplitude between a Dirichlet state $|D(x)\rrangle$ and a Neumann state $|N(y)\rrangle$.  Because there is no overlap between nonzero momentum and winding states, this reduces to the amplitude between $|F(1)\rrangle$ and $|F(-1)\rrangle$ with $\ov{\mathcal{C}(0)}\mathcal{C}(\pi)=1/\sqrt{2}$.  Comparing the result to \eqref{eq:SumOfDeltas} with $c_n=\sqrt{2}$, we get an exact match, a strong check of our calculations.

The other special case that deserves close consideration is $\theta_2=\theta_1$, in which case the gaps shrink to zero width, and the density of states is given by
\be
\label{eq:EqualBoundaryDensity}
\rho(h)=\left\{\begin{matrix} \frac{2\sqrt{2}\left|\mathcal{C}(\t_{1})\right|^2}{\pi\sqrt{h}}\sqrt{\frac{1-\cos(4\pi\sqrt{h})}{\cos(4\pi\sqrt{h})-\cos(2\theta_1)}}K\lp-\frac{1-\cos(4\pi\sqrt{h})}{\cos(4\pi\sqrt{h})-\cos(2\theta_1)}\rp, & \frac{1}{4}n^2\le h<\frac{1}{4}\lp n+\frac{\theta_1}{\pi}\rp^2, \\ \frac{2\sqrt{2}\left|\mathcal{C}(\t_{1})\right|^2}{\pi\sqrt{h}}\sqrt{\frac{1-\cos(4\pi\sqrt{h})}{\cos(2\theta_1)-\cos(4\pi\sqrt{h})}}K\lp-\frac{1-\cos(2\theta_1)}{\cos(2\theta_1)-\cos(4\pi\sqrt{h})}\rp, & \frac{1}{4}\lp n+\frac{\theta_1}{\pi}\rp^2\le h<\frac{1}{4}\lp n+1-\frac{\theta_1}{\pi}\rp^2, \\ \frac{2\sqrt{2}\left|\mathcal{C}(\t_{1})\right|^2}{\pi\sqrt{h}}\sqrt{\frac{1-\cos(4\pi\sqrt{h})}{\cos(4\pi\sqrt{h})-\cos(2\theta_1)}}K\lp-\frac{1-\cos(4\pi\sqrt{h})}{\cos(4\pi\sqrt{h})-\cos(2\theta_1)}\rp, & \frac{1}{4}\lp n+1-\frac{\theta_1}{\pi}\rp^2\le h<\frac{1}{4}\lp n+1\rp^2. \end{matrix}\right.
\ee
In particular, note that as $h$ approaches zero, governed by the first region above with $n=0$, there is no gap and $\rho(h)$ in fact approaches a constant value given by the $\theta_2\rightarrow\theta_1$ limit of~\eqref{eq:ValueAtGap}, which is $\rho(0)=4\pi|\mathcal{C}(\theta_1)|^2/\sin(\theta_1)$.  Thus the $\theta_1=\theta_2$ spectrum has no gap above $h=0$ and the identity operator in this sector is not isolated from the continuum.

\begin{figure}
    \centering
    \includegraphics[scale=0.42]{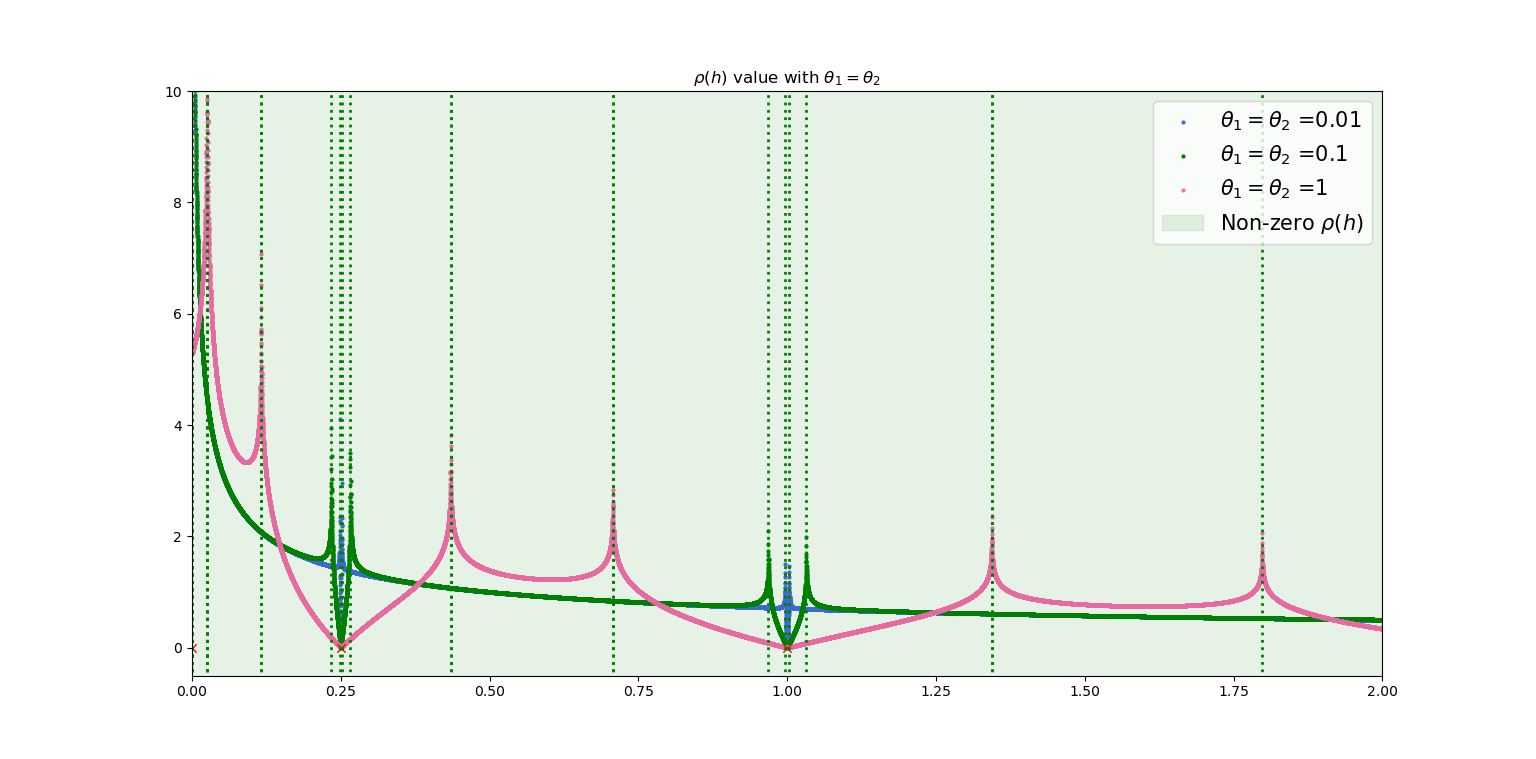}
    \caption{Three examples of the density of states when $\theta_1=\theta_2$.  As $\theta_1$ gets small, the spectrum approaches a continuous distribution with $\rho(h)\propto\sqrt{2/h}$ (although the divergent points persist for any finite value of $\theta_1$, if we subtract off the continuous piece and integrate $\rho$ over what remains, that quantity also vanishes in the limit).}
    \label{fig:DoSD}
\end{figure}

Three examples are plotted in Figure~\ref{fig:DoSD}.  When $\theta_1=\e\rr 0$, the middle region expands to fill most of the domain, and in the interior of this region we have a simple continuous density of states,
\be
\lim_{\e\rr 0}\rho(h)=\frac{\sqrt{2}\left|\mathcal{C}(0)\right|^2}{\sqrt{h}},\qquad h\ne\frac{1}{4}n^2.
\ee
On the other hand, when we are very near to points $n^2/4$, the density of states diverges in an integrable way.  Naively this would again mean that we have a sum of delta functions centered on squares of half-integers,

\be
\lim_{\e\rr 0}\rho(h)=|\mathcal{C}(0)|^{2} \lp \sqrt{\frac{2}{h}}+ \sum_{n=0}^{\infty}c_n\d\lp h-\frac{1}{4}n^2\rp\rp
\ee
However, actually the integrals of $\rho(h)$ around each of those points also vanish in the $\e\rr 0$ limit, so we would conclude that the coefficients of the delta functions are all vanishing, $c_n=0$.  This, in fact, matches well with our expectations. Friedan-Janik states don't include the $\theta=0$ case, but if we put $\theta=0$ in the expression, we get Ishibashi states with respect to the $U(1)$ current algebra, and these should satisfy
\begin{align}
    \left\llangle F(1)\left|q^H\right|F(1)\right\rrangle=\ & \left|\mathcal{C}(0)\right|^2\eta(q)^{-1}\non\\
    =\ & \left|\mathcal{C}(0)\right|^2\sqrt{\beta}\,\eta(\tilde{q})^{-1}\non\\
    =\ & \left|\mathcal{C}(0)\right|^2\int_0^\infty dh\,\sqrt{\frac{2}{h}}\frac{\tilde{q}^{\,h}}{\eta(\tilde{q})},
\end{align}
from which we read off the density of states in agreement with our result.
 \section{Pathologies of Friedan-Janik states}
\label{sec:Pathologies}

\subsection{Cluster condition violation}\label{cluster condition violation}

We will now show that if we use RCFT language, then we can deduce a contradiction between the results derived from Dirichlet/Neumann states and the results derived from the Friedan-Janik states. This RCFT language was also used in \cite{Janik:01} to derive the expression for the Friedan-Janik states. 

Before starting, we will write $||D(x_{0})\rrangle$ and $||N(\tilde{x}_{0})\rrangle$ in a convenient way. We note that the compact free boson theory has a $U(1)\times U(1)$ symmetry. Let's call the second $U(1)$ factor as $\widehat{U(1)}$. Now, the $||(N,0)\rrangle$ states preserve the $U(1)$ symmetry and $||(0,M)\rrangle$ states preserve the $\widehat{U(1)}$ symmetry. Let's denote the $||(N,0)\rrangle$ state for $N=0$ as $||(0,0)\rrangle_{U(1)}$ and the $||(0,M)\rrangle$ state for $M=0$ as $||(0,0)\rrangle_{\widehat{U(1)}}$. Both these states have to be a linear combination of $||[J, J]\rrangle$ states, since those are the states with $h=\bar{h}$ and are built from Virasoro primaries with zero momentum and winding.  When $J$ is even, every term in $|[J,J]\rangle$ is constructed by acting with an even number of $a^{\dagger}_{n}$ operators and an even number of $\widetilde{a}^{\dagger}_{n}$ operators, while when $J$ is odd, every term in $|[J,J]\rangle$ is constructed by acting with odd numbers of $a^{\dagger}_{n}$ and $\widetilde{a}^{\dagger}_{n}$ operators. This can be seen from the fact that at the self-dual radius the primary states are given by acting with $SU(2)$ lowering operators and the familiar fact that the $Y_{\ell 0}$ pick up $(-1)^\ell$ under a rotation by $\pi$ in the $xz$-plane. The same will then also be true of the Ishibashi states since every Virasoro raising operator involves even numbers of left- or right-moving operators. For example, for $J=1$ and $J=2$, we have the following states, up to a normalization factor, that we will set shortly,
\be
|[1,1]\rangle\propto a^{\dagger}_{1}\tilde{a}^{\dagger}_{1}|0\rangle,
\ee
\be
|[2,2]\rangle\propto\lp \frac{4}{\sqrt{3}}a^{\dagger}_{3}a^{\dagger}_{1}-2a^{\dagger}_{2}a^{\dagger}_{2}-\frac{2}{3}a^{\dagger}_{1}a^{\dagger}_{1}a^{\dagger}_{1}a^{\dagger}_{1}\rp\lp \frac{4}{\sqrt{3}}\tilde{a}^{\dagger}_{3}\tilde{a}^{\dagger}_{1}-2\tilde{a}^{\dagger}_{2}\tilde{a}^{\dagger}_{2}-\frac{2}{3}\tilde{a}^{\dagger}_{1}\tilde{a}^{\dagger}_{1}\tilde{a}^{\dagger}_{1}\tilde{a}^{\dagger}_{1}\rp|0\rangle.
\ee
The $||D(x_{0})\rrangle$ state preserves the $U(1)$ symmetry and thus, includes the $||(0,0)\rrangle_{U(1)}$ state. Similarly, the $||N(\tilde{x}_{0})\rrangle$ state preserves the $\widehat{U(1)}$ symmetry and thus, includes the $||(0,0)\rrangle_{\widehat{U(1)}}$ state. Suppose that $||(0,0)\rrangle_{U(1)}$ is given as 
\be\label{form of 0,0 U(1) with Ns}
||(0,0)\rrangle_{U(1)}=\sum_{J=0}^{\infty}N_{J}||[J,J]\rrangle.
\ee
This would imply
$$
\vphantom{\llangle (0,0)|q^{H}|(0,0)\rrangle}_{U(1)}\llangle (0,0)|q^{H}|(0,0)\rrangle_{U(1)}=\sum_{J,J'=0}^{\infty}\overline{N}_{J}N_{J'}\llangle [J,J]|q^{H}|[J',J']\rrangle
$$
$$
\Rightarrow \frac{1}{\eta(q)}=\sum_{J=0}^{\infty}\left|{N}_{J}\right|^{2}\frac{q^{J^{2}}-q^{(J+1)^{2}}}{\eta(q)}
$$
\be\label{N,N coefficient matching}
\Rightarrow 1=|N_{0}|^{2}+(|N_{1}|^{2}-|N_{0}|^{2})q+(|N_{2}|^{2}-|N_{1}|^{2})q^{4}+...\,.
\ee
By matching the coefficients on both sides, we conclude that;
\be\label{N coeff mods}
|N_{J}|^{2}=1\text{  for  }J=0,1,2,...\,.
\ee

Similarly, we can write down the expansion of $||(0,0)\rrangle_{\widehat{U(1)}}$ with some coefficients, say $M_{J}$ and conclude that $|M^{2}_{J}|=1$ for all $J$. In addition, we can use the overlap of a Dirichlet and Neumann boundary
\be\label{DN Cardy}
\llangle D(x_{0})|q^{H}|N(\tilde{x}_{0})\rrangle=\sqrt{\frac{\eta(q)}{\vartheta_{2}(q)}}
\ee
to get the following;
$$
\frac{1}{\sqrt{2}}\vphantom{\llangle (0,0)|q^{H}|(0,0)\rrangle}_{U(1)}\llangle (0,0)|q^{H}|(0,0)\rrangle_{\widehat{U(1)}}=\sqrt{\frac{\eta(q)}{\vartheta_{2}(q)}}
$$
$$
\Rightarrow \sum_{J=0}^{\infty}\overline{N}_{J}M_{J}\lp q^{J^{2}}-q^{(J+1)^{2}}\rp=\prod_{m=1}^{\infty}\frac{1-q^{m}}{1+q^{m}}
$$
\be\label{Dn expansion}
\Rightarrow \overline{N}_{0}M_{0}+(\overline{N}_{1}M_{1}-\overline{N}_{0}M_{0})q+(\overline{N}_{2}M_{2}-\overline{N}_{1}M_{1})q^{4}+...=1-2q+2q^{4}-2q^{9}+...\,,
\ee
where we used the identity
$$
\prod_{m=1}^{\infty}\frac{1-q^{m}}{1+q^{m}}=1+2\sum_{J=1}^{\infty}(-1)^{J}q^{J^{2}}.
$$
Now, comparing the coefficients in \eqref{Dn expansion}, we get the following;
\be\label{N,M relative normalization}
\overline{N}_{J}M_{J}=(-1)^{J}\text{  for  }J=0,1,2,...
\ee
We will choose the normalization such that $N_{J}=1$ for all $J$, which implies (because of \eqref{N,M relative normalization}) that $M_{J}=(-1)^{J}$. This is also consistent with the computations performed in~\cite{Janik:01}.  Therefore, we have the following expansions
\be\label{0,0 to JJ}
||(0,0)\rrangle_{U(1)}=\sum_{J=0}^{\infty}||[J,J]\rrangle,\efill ||(0,0)\rrangle_{\widehat{U(1)}}=\sum_{J=0}^{\infty}(-1)^{J}||[J,J]\rrangle.
\ee
Moreover, we define the following set of Ishibashi states, as they will lead us to real coefficients for Dirichlet and Neumann boundary states,
\begin{align}\label{diagonal n,m basis 1}
||(N,0)_{+}\rrangle=\frac{1}{\sqrt{2}}\left(||(N,0)\rrangle+||(-N,0)\rrangle\right),\;\;N>0,\\\label{diagonal n,m basis 2}
||(N,0)_{-}\rrangle=\frac{1}{i\sqrt{2}}\left(||(N,0)\rrangle-||(-N,0)\rrangle\right),\;\;N>0,\\\label{diagonal n,m basis 3}
||(0,M)_{+}\rrangle=\frac{1}{\sqrt{2}}\left(||(0,M)\rrangle+||(0,-M)\rrangle\right),\;\;M>0,\\\label{diagonal n,m basis 4}
||(0,M)_{-}\rrangle=\frac{1}{i\sqrt{2}}\left(||(0,M)\rrangle-||(0,-M)\rrangle\right),\;\;M>0.
\end{align}

Using equations \eqref{0,0 to JJ}-\eqref{diagonal n,m basis 4}, we can write \eqref{DirichletinIshibashi} as
$$
||D(x_{0})\rrangle=\frac{1}{\sqrt{\sqrt{2}R}}\left[\sum_{J=0}^{\infty}||[J,J]\rrangle+\sum_{N=1}^{\infty}\left(e^{\frac{iN}{R}x_{0}}||(N,0)\rrangle+e^{-\frac{iN}{R}x_{0}}||(-N,0)\rrangle\right)\right]
$$
\be\label{Dirichlet in +-}
=\frac{1}{\sqrt{\sqrt{2}R}}\left[\sum_{J=0}^{\infty}||[J,J]\rrangle+\sqrt{2}\sum_{N=1}^{\infty}\left[\cos{\left(\frac{Nx_{0}}{R}\right)}||(N,0)_{+}\rrangle-\sin{\left(\frac{Nx_{0}}{R}\right)}||(N,0)_{-}\rrangle\right]\right].
\ee
Similarly, \eqref{NeumanninIshibashi} becomes
\begin{multline}
\label{Neumann in +-}
||N(\tilde{x}_{0})\rrangle=\sqrt{\frac{R}{\sqrt{2}}}\left[\sum_{J=0}^{\infty}(-1)^{J}||[J,J]\rrangle\right.\\
\left. +\sqrt{2}\sum_{M=1}^{\infty}\left[\cos{\left(M\tilde{x}_{0}R\right)}||(0,M)_{+}\rrangle-\sin{\left(M\tilde{x}_{0}R\right)}||(0,M)_{-}\rrangle\right]\right].
\end{multline}
Now, we will show that if Friedan-Janik states \eqref{Friedanstate} satisfy the cluster condition, then they should produce a contradiction for the values of $M_{ij}^{\;\;k}$'s (which are defined in \eqref{m coefficients def}) deduced from $||D(x_{0})\rrangle$ and $||N(\tilde{x}_{0})\rrangle$. Before doing that, we need to consider different pairs of primary operators (say $i$ and $j$) defined in \eqref{diagonal n,m basis 1}-\eqref{diagonal n,m basis 4} and list all the primary operators (say $k$) such that the OPE coefficient $C_{ij}^{\;\;k}\neq 0$. Let's denote the operators corresponding to the Ishibashi states $||(N,0)\rrangle_{\pm},||(0,M)\rrangle_{\pm}$ and $||[J,J]\rrangle$ as $(N,0)_{\pm},(0,M)_{\pm},$ and $[J,J]$ respectively. The pairings involving just the $(N,0)_{\pm}$ operators give us the following
\begin{align}\label{(N,0) fusion rule 1}
(N,0)_{\pm}.(N',0)_{\pm}&: (N'-N,0)_{+}\text{ and }(N'+N,0)_{+}\text{ where }0<N<N'\\
\label{(N,0) fusion rule 2}
(N,0)_{+}.(N',0)_{-}&: (N'+N,0)_{-}\text{ and }(|N-N'|,0)_{-} \text{ where }N\neq N'\\
\label{(N,0) fusion rule 3}
(N,0)_{\pm}.(N,0)_{\pm}&: (2N,0)_{+}\text{ and }[J,J']\text{ where }J+J'\text{ is even}\\
\label{(N,0) fusion rule 4}
(N,0)_{+}.(N,0)_{-}&: (2N,0)_{-}\text{ and }[J,J']\text{ where }J+J'\text{ is odd}\\
\label{(N,0) fusion rule 5}
(N,0)_{\pm}.[J,J]&: (N,0)_{\pm}
\end{align}
Similarly, the pairings involving just the $(0,M)_{\pm}$ operators give us the following
\begin{align}\label{(0,M) fusion rule 1}
(0,M)_{\pm}.(0,M')_{\pm}&: (0,M'-M)_{+}\text{ and }(0,M'+M)_{+}\text{ where }0<M<M'\\
\label{(0,M) fusion rule 2}
(0,M)_{+}.(0,M')_{-}&: (0,M'+M)_{-}\text{ and }(0,|M-M'|)_{-}\text{ where }M\neq M'\\
\label{(0,M) fusion rule 3}
(0,M)_{\pm}.(0,M)_{\pm}&:(0,2M)_{+}\text{ and } [J,J']\text{ where }J+J'\text{ is even}\\
\label{(0,M) fusion rule 4}
(0,M)_{+}.(0,M)_{-}&: (0,2M)_{-}\text{ and } [J,J']\text{ where }J+J'\text{ is odd}\\
\label{(0,M) fusion rule 5}
(0,M)_{\pm}.[J,J]&:(0,M)_{\pm}
\end{align}
Lastly, the pairings between $(N,0)_{\pm}$ and $(0,M)_{\pm}$ give us the following
\begin{align}
(N,0)_{+}.(0,M)_{\pm}&: (N,M),(N,-M),(-N,M),\text{ and }(-N,-M)\\
(N,0)_{-}.(0,M)_{\pm}&: (N,M),(N,-M),(-N,M), \text{ and }(-N,-M)
\end{align}
Using the above listings, and the cluster condition \eqref{cluster condition}, we can deduce some of the values of and the relations between $M_{ij}^{\;\;k}$ coefficients. These calculations are given in appendix \ref{appendix A}. Using the results in appendix \ref{appendix A}, one can find a number of similar contradictions. A simple example of such contradictions can be built using \eqref{dirichlet n+n+},\eqref{neumann n+n+}, and \eqref{Friedan n+n+}. Using these equations, we can define a function as follows;
$$
f(\cos\t)=\sum_{J=0}^{\infty}M_{(N,0)_{+}(N,0)_{+}}^{[J,J]}P_{J}(\cos\t)=\begin{cases}
1\;\;\;\;\;\cos\t=1\\
0\;\;\;-1<\cos\t<1\\
0\;\;\;\;\;\cos\t=-1\\
\end{cases}
$$
which implies the following
$$
M_{(N,0)_{+}(N,0)_{+}}^{[J,J]}=\frac{2J+1}{2}\int_{-1}^{1} d\cos\t\;\; f(\cos \t)P_{J}(\cos\t)=0
$$
We see that this result is at odds with \eqref{dirichlet n+n+} and thus, we have a contradiction. So, using RCFT techniques, we have an argument that \eqref{Friedanstate} doesn't satisfy \eqref{cluster for RCFT} for $i$ and $j$ being non-zero momentum or winding operators.

In hindsight, the failure of the cluster condition should not be a surprise.  The derivation of the cluster condition from the Cardy-Lewellen sewing conditions depends on the presence of a gap above the vacuum in the open string sector.  As we showed in~\eqref{eq:EqualBoundaryDensity}, there is no gap in the $0<\theta_1=\theta_2<\pi$ spectrum, so there's no a priori reason that the cluster condition should hold here.  On the other hand, the cluster condition (applied only to the $[J,J]$ states) was used in~\cite{Janik:01} to derive the form of the Friedan-Janik states (which can also be derived by, for instance, taking a limit of the states constructed in~\cite{GaberdielRecknagel:2001}).  It may be that the application of the cluster condition can be justified in that case because it is only being applied to degenerate operators, whose correlation functions will obey additional constraints.  What we have shown here is that naively applying the cluster condition to non-degenerate states as well leads to a contradiction, but of course this is not a strong argument against the viability of the Friedan-Janik boundary states in general.  It would be interesting to go back to the full sewing conditions and attempt to perform a more complete consistency analysis, or else derive a modified version of the cluster condition as is done, for example, in~\cite{Schomerus:2005aq} for Liouville theory.

\section{Conclusions and future directions}\label{sec:Conclusions}
In this article we have undertaken a more detailed study of the Friedan-Janik boundary states $|[J,J]\rrangle$.  Open string sectors between these boundary states generically have continuous spectra, and we were able to give an explicit expression for the density of states in every such sector, at least up to a normalization factor (which is suggested to be infinite by showing that the boundary entropy of the Friedan-Janik boundary states is infinite\cite{Tseng:2002ax}).

Besides the continuous spectrum of states, these boundary states exhibit certain other pathologies.  They don't correspond in any simple way to a boundary condition relating the antiholomorphic part of the boson field to the holomorphic part.  One can argue for a failure of the cluster condition arising from the continuum of intermediate states which can appear in the two-point function in the presence of the boundary. And finally, and most quantifiably, the $g$ function of these boundary states diverges~\cite{Tseng:2002ax}, indicating the presence of an infinite number of degrees of freedom at the boundary.

As a consequence of the divergence of the $g$-function, we do not expect that these should arise spontaneously in a physical system.  Since $g$ decreases monotonically under boundary RG flows \cite{Affleck:1991tk,Friedan:2003yc}, we can not hope to obtain FJ states from a boundary perturbation of a Neumann or Dirichlet state, or even from a Neumann or Dirichlet state dressed with finitely many additional degrees of freedom (e.g.~Chan-Paton factors), however it is an interesting open question whether these states could arise as the end-point of a bulk RG flow from a boundary conformal field theory with $c>1$, since it is known that $g$ can increase under such flows \cite{Green:2007wr}.

Setting aside the issue of how one might engineer a theory with such a boundary state, it is also interesting to discuss the fate of such a boundary state if it is present initially.  Are these states unstable?  There are no obvious perturbative instabilities, but one suspects that there may be non-perturbative mechanisms which may engage.  Indeed, since one interpretation of the FJ states is as a smearing of an infinite number of Neumann or Dirichlet states, previous works~\cite{Tong:2002rq,Harvey:2005ab} suggest that worldsheet instanton effects might play a role, perhaps localizing the state to a finite combination of Neumann or Dirichlet boundary states.

Finally, it would be very intriguing to repeat this sort of analysis in certain other contexts, primarily of multiple bosons (Narain CFTs), orbifolds of these theories, or more generally in non-linear sigma model CFTs.  In the latter case one might be able to use exact descriptions such as orbifolds or Gepner models to hunt for analogous boundary states.  We hope to turn to such efforts in the future.

\section*{Acknowledgements}

The authors would like to thank J.~Distler, M.~Gaberdiel, R.~Janik, Z.~Komargodski, and O.~Lunin for useful discussions.
\newpage
\appendix
\section{Constraints on $M_{ij}^{k}$ coefficients}\label{appendix A}
The indices $i$ and $j$ in the cluster condition \eqref{cluster condition} can take several different values but we won't get an independent equation if we switched $i\leftrightarrow j$ because $M_{ij}^{k}$ coefficients contain the OPE coefficients $C^{k}_{ij}$ which are symmetric in $i$ and $j$. We can consider all the $(i,j)$ tuples for the Dirichlet, Neumann, and Friedan-Janik boundary conditions that will give a nontrivial condition. That's what we do in the following.
\subsection*{Dirichlet boundary}
Using \eqref{Dirichlet in +-}, we can read off the $B_{\alpha i}$ coefficients for the Dirichlet state to be the following
\be\label{B Dirichlet}
B_{D(x_{0})(N,0)_{+}}=\sqrt{2}\cos{\left(\frac{Nx_{0}}{R}\right)}\efill B_{D(x_{0})(N,0)_{-}}=-\sqrt{2}\sin{\left(\frac{Nx_{0}}{R}\right)}\efill B_{D(x_{0})[J,J]}=1
\ee
with other $B_{\al i}$ being zero. The relations between the relevant $M_{ij}^{k}$ coefficients are calculated in the following;
\begin{itemize}
\item$i=j=(N,0)_{+}:$
$$
B_{D(x_{0})(N,0)_{+}}B_{D(x_{0})(N,0)_{+}}=M_{(N,0)_{+}(N,0)_{+}}^{(2N,0)_{+}}B_{D(x_{0})(2N,0)_{+}}+\sum_{J=0}^{\infty}M_{(N,0)_{+}(N,0)_{+}}^{[J,J]}B_{D(x_{0})[J,J]}
$$
$$
\Rightarrow 2\cos^{2}{\left(\frac{Nx_{0}}{R}\right)}=\sqrt{2}M_{(N,0)_{+}(N,0)_{+}}^{\;\;(2N,0)_{+}}\cos\left(\frac{2Nx_{0}}{R}\right)+\sum_{J=0}^{\infty}M_{(N,0)_{+}(N,0)_{+}}^{[J,J]}
$$
\be\label{dirichlet n+n+}
\Rightarrow M_{(N,0)_{+}(N,0)_{+}}^{(2N,0)_{+}}=\frac{1}{\sqrt{2}}\efill \sum_{J=0}^{\infty}M_{(N,0)_{+}(N,0)_{+}}^{[J,J]}=1
\ee
\item$i=(N,0)_{+},\;j=(N,0)_{-}:$
$$
-2\cos{\left(\frac{Nx_{0}}{R}\right)}\sin{\left(\frac{Nx_{0}}{R}\right)}=-\sqrt{2}M_{(N,0)_{+}(N,0)_{-}}^{\;\;(2N,0)_{-}}\sin\left(\frac{2Nx_{0}}{R}\right)
$$
\be\label{dirichlet n+n-}
\Rightarrow M_{(N,0)_{+}(N,0)_{-}}^{(2N,0)_{-}}=\frac{1}{\sqrt{2}}
\ee
\item$i=(N,0)_{+},\;j=(N',0)_{+} \text{ with }N< N':$
$$
\sqrt{2}\cos{\left(\frac{Nx_{0}}{R}\right)}\cos{\left(\frac{N'x_{0}}{R}\right)}=M_{(N,0)_{+}(N',0)_{+}}^{\;\;(N+N',0)_{+}}\cos{\left(\frac{(N+N')x_{0}}{R}\right)}+M_{(N,0)_{+}(N',0)_{+}}^{\;\;(N'-N,0)_{+}}\cos{\left(\frac{(N'-N)x_{0}}{R}\right)}
$$
$$
=(M_{(N,0)_{+}(N',0)_{+}}^{\;\;(N+N',0)_{+}}+M_{(N,0)_{+}(N',0)_{+}}^{\;\;(N'-N,0)_{+}})\cos{\left(\frac{Nx_{0}}{R}\right)}\cos{\left(\frac{N'x_{0}}{R}\right)}
$$
$$
+(M_{(N,0)_{+}(N',0)_{+}}^{\;\;(N'-N,0)_{+}}-M_{(N,0)_{+}(N',0)_{+}}^{\;\;(N+N',0)_{+}})\sin{\left(\frac{Nx_{0}}{R}\right)}\sin{\left(\frac{N'x_{0}}{R}\right)}
$$
\be\label{dirichlet n+n'+}
\Rightarrow M_{(N,0)_{+}(N',0)_{+}}^{\;\;(N'-N,0)_{+}}=M_{(N,0)_{+}(N',0)_{+}}^{\;\;(N+N',0)_{+}}=\frac{1}{\sqrt{2}}.
\ee
\item$i=(N,0)_{+},\;j=(N',0)_{-} \text{ with }N> N':$
$$
\sqrt{2}\cos{\left(\frac{Nx_{0}}{R}\right)}\sin{\left(\frac{N'x_{0}}{R}\right)}=M_{(N,0)_{+}(N',0)_{-}}^{\;\;(N+N',0)_{-}}\sin{\left(\frac{(N+N')x_{0}}{R}\right)}+M_{(N,0)_{+}(N',0)_{-}}^{\;\;(N-N',0)_{-}}\sin{\left(\frac{(N-N')x_{0}}{R}\right)}
$$
\be\label{dirichlet n+n'-}
\Rightarrow M_{(N,0)_{+}(N',0)_{-}}^{(N+N',0)_{-}}=-M_{(N,0)_{+}(N',0)_{-}}^{(N-N',0)_{-}}=\frac{1}{\sqrt{2}}
\ee
\item$i=(N,0)_{+},\;j=[J,J]:$
\be\label{dirichlet n+J}
\sqrt{2}\cos\lp\frac{Nx_{0}}{R}\rp=M_{(N,0)_{+}[J,J]}^{(N,0)_{+}}\sqrt{2}\cos\lp\frac{Nx_{0}}{R}\rp\Rightarrow M_{(N,0)_{+}[J,J]}^{(N,0)_{+}}=1
\ee
\item$i=(N,0)_{-},\;j=(N,0)_{-}:$
$$
2\sin^{2}{\left(\frac{Nx_{0}}{R}\right)}=\sqrt{2}M_{(N,0)_{-}(N,0)_{-}}^{\;\;(2N,0)_{+}}\cos\left(\frac{2Nx_{0}}{R}\right)+\sum_{J=0}^{\infty}M_{(N,0)_{-}(N,0)_{-}}^{[J,J]}
$$
\be\label{dirichlet n-n-}
\Rightarrow M_{(N,0)_{-}(N,0)_{-}}^{(2N,0)_{+}}=-\frac{1}{\sqrt{2}}\efill \sum_{J=0}^{\infty}M_{(N,0)_{-}(N,0)_{-}}^{[J,J]}=1
\ee
\item$i=(N,0)_{-},\;j=(N',0)_{-}\text{ with }N<N':$
\begin{align*}
\sqrt{2}\sin{\left(\frac{Nx_{0}}{R}\right)}\sin{\left(\frac{N'x_{0}}{R}\right)}=M_{(N,0)_{-}(N',0)_{-}}^{\;\;(N+N',0)_{+}}\cos{\left(\frac{(N+N')x_{0}}{R}\right)}\\
+M_{(N,0)_{-}(N',0)_{-}}^{\;\;(N'-N,0)_{+}}\cos{\left(\frac{(N'-N)x_{0}}{R}\right)}
\end{align*}
\be\label{dirichlet n-n'-}
\Rightarrow M_{(N,0)_{-}(N',0)_{-}}^{(N+N')_{-}}=-M_{(N,0)_{-}(N',0)_{-}}^{(N'-N)_{-}}=-\frac{1}{\sqrt{2}}
\ee
\item$i=(N,0)_{-},\;j=[J,J]:$
\be\label{dirichlet n-J}
-\sqrt{2}\sin\lp\frac{Nx_{0}}{R}\rp=-\sqrt{2}M_{(N,0)_{-}[J,J]}^{(N,0)_{-}}\sin\lp\frac{Nx_{0}}{R}\rp\Rightarrow M_{(N,0)_{-}[J,J]}^{(N,0)_{-}}=1
\ee
\item$i=(0,M)_{+},\;j=(0,M)_{+}:$
\be\label{dirichlet m+m+}
\sum_{J=0}^{\infty}M_{(0,M)_{+}(0,M)_{+}}^{[J,J]}=0
\ee
\item$i=[J,J],\;j=[J',J']\text{ with }J\leq J':$
\be\label{dirichlet JJ'}
\sum_{k=0}^{J}M_{[J,J][J',J']}^{[J'-J+2k,J'-J+2k]}=1
\ee
\end{itemize}
The rest of the cases give us trivial equations.
\subsection*{Neumann boundary}
Using \eqref{Neumann in +-}, we can read off the $B_{\alpha i}$ coefficients for the Neumann state to be the following
\be\label{B Neumann}
B_{N(\tilde{x}_{0})(0,M)_{+}}=\sqrt{2}\cos{\left(M\tilde{x}_{0}R\right)}\efill B_{N(\tilde{x}_{0})(0,M)_{-}}=-\sqrt{2}\sin{\left(M\tilde{x}_{0}R\right)}\efill B_{N(\tilde{x}_{0})[J,J]}=(-1)^{J}
\ee
with other $B_{\al i}$'s being zero. The relations between the relevant $M_{ij}^{k}$ coefficients are as follows;
\begin{itemize}
\item$i=j=(N,0)_{+}:$
\be\label{neumann n+n+}
\sum_{J=0}^{\infty}(-1)^{J}M^{[J,J]}_{(N,0)_{+}(N,0)_{+}}=0
\ee
\item$i=(N,0)_{-},\;j=(N,0)_{-}:$
\be\label{neumann n-n-}
\sum_{J=0}^{\infty}(-1)^{J}M^{[J,J]}_{(N,0)_{+}(N,0)_{+}}=0
\ee
\item$i=(0,M)_{+},\;j=(0,M)_{+}:$
$$
2\cos^{2}(M\tilde{x}_{0}R)=\sqrt{2}\cos(2M\tilde{x}_{0}R)M_{(0,M)_{+}(0,M)_{+}}^{(0,2M)_{+}}+\sum_{J=0}^{\infty}(-1)^{J}M_{(0,M)_{+}(0,M)_{+}}^{[J,J]}
$$
\be\label{neumann m+m+}
M_{(0,M)_{+}(0,M)_{+}}^{(0,2M)_{+}}=\frac{1}{\sqrt{2}}\efill \efill \sum_{J=0}^{\infty}(-1)^{J}M_{(0,M)_{+}(0,M)_{+}}^{[J,J]}=1
\ee
\item$i=(0,M)_{+},\;j=(0,M)_{-}:$
$$
-2\cos(M\tilde{x}_{0}R)\sin(M\tilde{x}_{0}R)=\sqrt{2}M_{(0,M)_{+}(0,M)_{-}}^{(0,2M)_{-}}\sin(2M\tilde{x}_{0}R)
$$
\be\label{neumann m+m-}
\Rightarrow M_{(0,M)_{+}(0,M)_{-}}^{(0,2M)_{-}}=-\frac{1}{\sqrt{2}}
\ee
\item$i=(0,M)_{+},\;j=(0,M')_{+}\text{ with }M< M':$
\begin{align*}
2\cos(M\tilde{x}_{0}R)\cos(M'\tilde{x}_{0}R)=\sqrt{2} M_{(0,M)_{+}(0,M')_{+}}^{(0,M'-M)_{+}}\cos((M'-M)\tilde{x}_{0}R)\\
+\sqrt{2}M_{(0,M)_{+}(0,M')_{+}}^{(0,M'-M)_{+}}\cos((M'-M)\tilde{x}_{0}R)
\end{align*}
\be\label{neumann m+m'+}
\Rightarrow M^{(0,M'+M)_{+}}_{(0,M)_{+}(0,M')_{+}}=M^{(0,M'-M)_{+}}_{(0,M)_{+}(0,M')_{+}}=\frac{1}{\sqrt{2}}
\ee
\item$i=(0,M)_{+},\;j=(0,M')_{-}\text{ with }M< M':$
\begin{align*}
\sqrt{2}\cos(M\tilde{x}_{0}R)\sin(M'\tilde{x}_{0}R)=M_{(0,M)_{+}(0,M')_{-}}^{(0,M+M')}\sin((M+M')\tilde{x}_{0}R)\\
+M_{(0,M)_{+}(0,M')_{-}}^{(0,M'-M)}\sin((M'-M)\tilde{x}_{0}R)
\end{align*}
$$
\Rightarrow M^{(0,M+M')_{-}}_{(0,M)_{+}(0,M')_{-}}=M^{(0,M'-M)_{-}}_{(0,M)_{+}(0,M')_{-}}=\frac{1}{\sqrt{2}}
$$
\item$i=(0,M)_{+},\;j=(0,M')_{-}\text{ with }M> M':$
\begin{align*}
\sqrt{2}\cos(M\tilde{x}_{0}R)\sin(M'\tilde{x}_{0}R)=M_{(0,M)_{+}(0,M')_{-}}^{(0,M+M')}\sin((M+M')\tilde{x}_{0}R)\\
+M_{(0,M)_{+}(0,M')_{-}}^{(0,M-M')}\sin((M-M')\tilde{x}_{0}R)
\end{align*}
$$
\Rightarrow M^{(0,M+M')_{-}}_{(0,M)_{+}(0,M')_{-}}=-M^{(0,M-M')_{-}}_{(0,M)_{+}(0,M')_{-}}=\frac{1}{\sqrt{2}}
$$

\item$i=(0,M)_{+},\;j=[J,J]:$
\be\label{neumann m+J}
\sqrt{2}\cos(M\tilde{x}_{0}R)=\sqrt{2}M_{(0,M)_{+}[J,J]}^{(0,M)_{+}}\cos(M\tilde{x}_{0}R)\Rightarrow M_{(0,M)_{+}[J,J]}^{(0,M)_{+}}=1
\ee
\item$i=(0,M)_{-},\;j=(0,M)_{-}:$
$$
2\sin^{2}(M\tilde{x}_{0}R)=\sqrt{2}M^{(0,2M)_{+}}_{(0,M)_{-}(0,M)_{-}}\cos(2M\tilde{x}_{0}R)+\sum_{J=0}^{\infty}(-1)^{J}M_{(0,M)_{-}(0,M)_{-}}^{[J,J]}
$$
\be\label{neumann m-m-}
\Rightarrow M^{(0,2M)_{+}}_{(0,M)_{-}(0,M)_{-}}=-\frac{1}{\sqrt{2}}\efill\efill \sum_{J=0}^{\infty}(-1)^{J}M_{(0,M)_{-}(0,M)_{-}}^{[J,J]}=1
\ee
\item$i=(0,M)_{-},\;j=(0,M')_{-}\text{ with } M< M':$
\begin{align*}
2\sin(M\tilde{x}_{0}R)\sin(M'\tilde{x}_{0}R)=\sqrt{2}M^{(0,M'-M)_{+}}_{(0,M)_{-}(0,M')_{-}}\cos((M'-M)\tilde{x}_{0}R)\\
+\sqrt{2}M^{(0,M'+M)_{+}}_{(0,M)_{-}(0,M')_{-}}\cos((M'+M)\tilde{x}_{0}R)
\end{align*}
\be\label{neumann m-m'-}
\Rightarrow M^{(0,M'+M)_{+}}_{(0,M)_{-}(0,M')_{-}}=-M^{(0,M'-M)_{+}}_{(0,M)_{-}(0,M')_{-}}=-\frac{1}{\sqrt{2}}
\ee
\item$i=(0,M)_{-},\;j=[J,J]:$
\be\label{neumann m-J}
-\sqrt{2}\sin(M\tilde{x}_{0}R)=-\sqrt{2}M_{(0,M)_{-}[J,J]}^{(0,M)_{-}}\sin(M\tilde{x}_{0}R)\Rightarrow M_{(0,M)_{-}[J,J]}^{(0,M)_{-}}=1
\ee
Other cases either give trivial equations, or give equations that we have already encountered.
\end{itemize}
\subsection*{Friedan-Janik boundary ($-1< \cos\theta<1$)}
\begin{itemize}
\item$i=j=(N,0)_{+}:$
\be\label{Friedan n+n+}
\sum_{J=0}^{\infty}M_{(N,0)_{+}(N,0)_{+}}^{[J,J]}P_{J}(\cos\theta)=0
\ee
\item$i=(N,0)_{-},\;j=(N,0)_{-}:$
\be\label{Friedan n-n-}
\sum_{J=0}^{\infty}M_{(N,0)_{-}(N,0)_{-}}^{[J,J]}P_{J}(\cos\theta)=0
\ee
\item$i=(0,M)_{+},\;j=(0,M)_{+}:$
\be\label{Friedan m+m+}
\sum_{J=0}^{\infty}M_{(0,M)_{+}(0,M)_{+}}^{[J,J]}P_{J}(\cos{\theta})=0
\ee
\item$i=(0,M)_{-},\;j=(0,M)_{-}:$
\be\label{Friedan m-m-}
\sum_{J=0}^{\infty}M_{(0,M)_{-}(0,M)_{-}}^{[J,J]}P_{J}(\cos{\theta})=0
\ee
\item$i=[J,J],\;j=[J',J']\text{ with }J\leq J':$
\be\label{Friedan JJ}
P_{J}(\cos{\theta})P_{J'}(\cos{\theta})=\sum_{k=0}^{J}M_{[J,J][J',J']}^{[J'-J+2k,J'-J+2k]}P_{J'-J+2k}(\cos\theta)
\ee
\end{itemize}
The rest of the cases give us trivial equations.
\bibliographystyle{apsrev4-2}
\bibliography{References}
\end{document}